\documentclass[manuscript,showpacs,preprintnumbers,amsmath,amssymb]{revtex4-1}
\usepackage{graphicx}
\usepackage{rotating}
\usepackage{amssymb}
\usepackage{amsfonts}
\usepackage{amsmath}

\newcommand{\be}{\begin{equation}}
\newcommand{\ee}{\end{equation}}
\newcommand{\bea}{\begin{eqnarray}}
\newcommand{\eea}{\end{eqnarray}}

\begin{document}

\title{Puzzles of the dark energy in the universe - phantom}

\author{Mariusz P. D\c{a}browski}
\email{mpdabfz@wmf.univ.szczecin.pl}
\affiliation{\it Institute of Physics, University of Szczecin, Wielkopolska 15, 70-451 Szczecin, Poland}
\affiliation{\it Copernicus Center for Interdisciplinary Studies,
S{\l }awkowska 17, 31-016 Krak\'ow, Poland}

\date{\today}

\input epsf

\begin{abstract}
This paper is devoted to some simple approach based on general physics tools to describe the physical properties of a hypothetical particle which can be the source of dark energy in the Universe known as phantom. Phantom is characterized by the fact that it possesses negative momentum and kinetic energy and that it gives large negative pressure which acts as antigravity. We consider phantom harmonic oscillator in comparison to a standard harmonic oscillator. By using the first law of thermodynamics we explain why the energy density of the Universe grows when it is filled with phantom. We also show how the collision of phantom with a standard particle leads to exploration of energy from the former by the latter (i.e. from phantom to the standard) if their masses are different. The most striking of our conclusions is that the collision of phantom and standard particles of the same masses is impossible unless both of them are at rest and suddenly start moving with the opposite velocities and kinetic energies. This effect is a classic analogue of a quantum mechanical particle pair creation in a strong electric field or in physical vacuum.
\end{abstract}

%Kew words: cosmology: dark energy; phantom: oscillators, energy, collisions


\maketitle

\section{Introduction. What is phantom?}
\label{intro}
\setcounter{equation}{0}

The large-scale structure and the evolution of the Universe is determined by gravity which is well approximated by Einstein's general relativity \cite{HE}. The fact that the Universe is expanding was known since the discovery of Hubble in 1929 \cite{hubble}, but only recently by measuring the flux from distant supernovae of type Ia, it was realized that the expansion of the universe is accelerating \cite{supernovae}. This can be expressed in terms of the scale factor $a(t)$ (the function which describes how distances in the Universe change in time) by the condition $\ddot{a} >0$, and related to general relativity as the violation of the strong energy condition $\varepsilon + 3p \geq 0$ ($\varepsilon$ - the energy density, $p$ - the pressure) which corresponds to the statement that gravity is to be an attractive force \cite{wald}. Phenomenologically, the violation means that there must be some kind of matter in the universe which has negative pressure acting as antigravity and dates us back to an early idea of Einstein, who introduced such a repulsive force under the name of the cosmological constant $\Lambda$ ($p_{\Lambda} = - \varepsilon_{\Lambda} = - \Lambda c^2/(8 \pi G)$, $c$ - speed of light, $G$ - gravitational constant) \cite{einstein}. Negative pressure, though sounds exotic, is known in common physics as capillarity inside the trees and was obtained in the laboratory by stretching some water in a vessel already in 1850 \cite{negpress}.

According to particle physics, the best physical interpretation for the cosmological constant is the quantum vacuum energy \cite{perkins}, but the field theory evaluation of its value is over 120 orders of magnitude higher than the value obtained by observations based on general relativity. This is why alternative proposals have been presented and are known under the name of dark energy \cite{shinji}. There is yet another issue. If the pressure of any candidate for dark energy is less than that of the cosmological constant, i.e. if $p \geq - \varepsilon$ which in the formalism of general relativity is known as the null energy condition \cite{HE}, then the so-called cosmic ``no-hair" theorem holds. It says that no matter what is the other content of the universe apart from even a tiny amount of the cosmological constant, then this constant energy will dominate the universe evolution in future (exhibiting obviously the negative pressure and leading to a totally diluted empty universe).

However, bearing in mind the observational constraints on the expansion of the universe coming from more recent data \cite{supernew}, it emerged that it was possible to have the dark energy in the Universe which violated the null energy condition, i.e. which had $p \leq -\varepsilon$. This matter which exhibits supernegative pressure was first proposed by Caldwell \cite{robert} and dubbed as phantom in reference to a popular film ``Star Wars''. At first glance phantom does not look very much harmful than just some small negative pressure matter, but after more careful insight, it emerges that it really brings lots of physical puzzles. First of all, phantom matter (or phantom particles) violate all the energy conditions of general relativity. This refers to already mentioned the strong and the null energy conditions as well as to the weak energy condition $\varepsilon + p \geq 0$, $\varepsilon \geq 0$ and the dominant energy condition $\mid p \mid \leq \varepsilon$, $\varepsilon \geq 0$). All this means that phantom has very unusual properties as considered in view of standard physics. In particular, it violates the before mentioned cosmic ``no hair'' theorem which means that even a tiny amount of phantom in the Universe may dominate its evolution instead of the cosmological constant \cite{CCW,PRD03}. Besides, it leads to both classical and quantum instabilities due to its negative kinetic energy and momentum \cite{cline,instab,fulvio}. However, in view of the observational constraints
phantom matter is one of the candidate for dark energy which drives the accelerated expansion of the universe and still cannot be rejected. This is why we think that it is useful to discuss some of its properties on the level of general physics classes, even for a reader who is not advanced in all the subtleties of contemporary particle physics and cosmology.

In this paper we will explore some of the effects which emerge for phantom on a basic physics level. In Section \ref{harmonic} we will discuss the properties of a classical phantom harmonic oscillator and show how it differs from a standard one. In Section \ref{thermo} we discuss the properties of phantom in the context of the first law of thermodynamics or the energy conservation. In Section \ref{collision} we will discuss the effect of an elastic classical collision of a phantom particle (of negative momentum and kinetic energy) with a standard particle (of positive momentum and kinetic energy). In Section \ref{summary} we will give the summary of our results.

\section{Phantom harmonic oscillator}
\label{harmonic}
\setcounter{equation}{0}

We start with the discussion of a phantom harmonic oscillator by a short discussion of a standard harmonic oscillator which has the total energy composed
of the kinetic $E_{ks}$ and potential $E_p$ energies as follows
\be
E = E_{ks} + E_p = \frac{1}{2} mv^2 + \frac{1}{2} kx^2~~,
\ee
where $m$ is the mass, $v$ the velocity, $k$ the elastic constant, and $x$ the displacement. As one can see, the potential energy is a quadratic function which has
a minimum at $x=0$ and so the particle is oscillating around this minimum between some values $- x_{max} < x < x_{max}$. Obviously, we have for some $x=x_1$ and $x=x_{max}$ that
\bea
E(x_1) &=& \frac{1}{2} m v_1^2 + \frac{1}{2} k x_1^2~~, \\
E(0) &=& \frac{1}{2} m v_0^2, \hspace{0.4cm} E(x_{max}) = \frac{1}{2} k x_{max}^2~~.
\eea
If the energy is conserved, then the larger is the displacement, the smaller is the velocity of a particle, i.e.
\be
\frac{1}{2} m \left( v_0^2 - v_1^2 \right) = \frac{1}{2} k x_1^2~~,
\ee
which means that
\be
v_0^2 > v_1^2, \hspace{0.2cm} {\rm or} \hspace{0.2cm} \mid v_0 \mid > \mid v_1 \mid~~.
\ee
Lagrangian of this standard system has a simple and well-known form
\be
L = \frac{1}{2} m \dot{x}^2 - \frac{1}{2} k x^2~~,
\ee
and so the Euler-Lagrange equation \cite{morin}
\be
\frac{\partial L}{\partial x} = \frac{\partial}{\partial t} \frac{\partial L}{\partial \dot{x}}
\ee
give just the Newton's equation
\be
-k x = m \ddot{x}~,
\ee
which has a periodic solution
\be
\label{oscill}
x(t) = A \cos{(\omega t + \varphi)}~~,
\ee
where $\omega = \sqrt{k/m}$ and $A, \varphi=$ const. This means the particle is oscillating around the equilibrium.

After this simple reminder about the standard harmonic oscillator we now consider the two examples of the phantom oscillators.

The first of them is characterized by the {\it negative} kinetic energy $E_{kf}$ and the same potential energy $E_p$ as in the standard case, i.e.
\be
E = E_{kf} + E_p = - \frac{1}{2} mv^2 + \frac{1}{2} k x^2~~.
\ee
Since now we have $E_{kf} < 0$, then
\bea
E(x_1) &=& -\frac{1}{2} m v_1^2 + \frac{1}{2} k x_1^2~~, \\
E(0) &=& -\frac{1}{2} m v_0^2, \hspace{0.4cm} E(x_{max}) = \frac{1}{2} k x_{max}^2~~,
\eea
so that assuming the energy conservation we have
\be
\frac{1}{2} m \left( v_1^2 - v_0^2 \right) = \frac{1}{2} k x_1^2~~,
\ee
and so
\be
v_1^2 > v_0^2, \hspace{0.2cm} {\rm or} \hspace{0.2cm} \mid v_1 \mid > \mid v_0 \mid~~,
\ee
which means that the (negative) velocity is increasing if the particle is going away from the equilibrium point placed at $x=0$ and can grow indefinitely. In other words, the (negative) kinetic energy of phantom becomes more and more negative, finally approaching minus infinity. In fact, the particle (phantom) can get infinite potential energy at the expense of the negative (and finally also indefinite) kinetic energy. This is an unstable state which can be noticed by analyzing the Lagrangian
\be
L = -\frac{1}{2} m \dot{x}^2 - \frac{1}{2} k x^2~~,
\ee
which shows that the canonical momentum
\be
p = \frac{\partial L}{\partial \dot{x}} = - m \dot{x}
\ee
is negative (since $x$ grows, i.e. $\dot{x} >0$ during the movement from $x=0$ to $x=x_1$). The Euler-Lagrange equation now reads as
\be
- kx = - m\ddot{x}~~,
\ee
and is solved by
\be
x = A \cosh{(\omega t + \varphi)}~~,
\ee
where $\omega = \sqrt{k/m}$. It tells us that the motion is not oscillatory and unbounded (which gives an instability)

The second phantom oscillator has negative kinetic energy $E_{kf}$ with an ``upside down'' potential $\bar{E}_p$ which has a maximum instead of a minimum and is still quadratic. Its
total energy reads as
\be
E = E_{kf} + \bar{E}_p = - \frac{1}{2} mv^2 - \frac{1}{2} k x^2~~.
\ee
The Lagrangian is
\be
L = -\frac{1}{2} m \dot{x}^2 + \frac{1}{2} k x^2~~,
\ee
and gives the equation of motion in the form
\be
\label{oscf2}
kx = - m\ddot{x}
\ee
also with the negative canonical momentum
\be
p = \frac{\partial L}{\partial \dot{x}} = - m \dot{x}~~.
\ee
However, the main point now is that the solution of (\ref{oscf2}) is harmonic (i.e. given by (\ref{oscill}) despite the fact that the particle at maximum displacement has the smallest potential energy). We can now write that
\bea
E(x_1) &=& -\frac{1}{2} m v_1^2 - \frac{1}{2} k x_1^2~~, \\
E(0) &=& -\frac{1}{2} m v_0^2, \hspace{0.4cm} E(x_{max}) = -\frac{1}{2} k x_{max}^2~~,
\eea
so that assuming the energy conservation, one gets
\be
\frac{1}{2} m \left( v_0^2 - v_1^2 \right) = \frac{1}{2} k x_1^2~~,
\ee
which means that
\be
v_0^2 > v_1^2, \hspace{0.2cm} {\rm or} \hspace{0.2cm} \mid v_0 \mid > \mid v_1 \mid~~.
\ee
The conclusion is that the kinetic energy at equilibrium $x=0$ is smaller than at the turning point $x_{max}$, i.e.
\be
- \frac{1}{2} m v_0^2 < - \frac{1}{2} m v_1^2 < - \frac{1}{2} m v_{max}^2
\ee
and obviously at $x=0$, the particle cannot be at rest (since $v_0 = 0$ would give $0 < - \frac{1}{2} m v_1^2$ which is a contradiction), while
it can be at rest for $x=x_{max}$ (since $v_{max} = 0$, gives $- \frac{1}{2} m v_0^2 < 0$ which is acceptable). In other words, the particle at an equilibrium point
has smaller (negative) kinetic energy than outside of this point, so it somehow ``falls'' into the minimum of the kinetic energy at $x=0$ instead of the minimum
of the potential at $x=x_{max}$.

\section{Thermodynamics, dark energy and phantom}
\label{thermo}
\setcounter{equation}{0}

According to the 1st law of thermodynamics which is just the statement of the energy conservation, a change of the internal energy of a system $dE$ is equal the sum
of the heat inflow into the system $dQ$ and the work done {\it on} the system $dW$ \cite{young}, i.e.
\be
\label{1stlaw}
dE = dQ + dW~.
\ee
One may also use the entropy $S$ and write down the heat as $dQ = TdS$, where $T$ is the temperature at which the heat is transferred.
We employ the thermodynamical description to the universe as a whole in an adiabatic approximation which means that we exclude a possibility to exchange the heat
(i.e. consider an isolated system with $dQ=0$) and this is equivalent to the statement that the entropy of the universe is constant $S=$ const., and its change is $dS = 0$. Then any change of the universe energy will be equal to the work done by the gravitational forces while changing the volume. If the universe expands $(dV > 0)$ - it does positive work $pdV$
(work is done {\it by the system}) which implies that the system's internal energy is diminishing (which is an analogy of the simple expansion of the gas which looses its
internal energy). This can be written down as
\be
\label{energy}
dE = - p dV~,
\ee
where $dV$ is the growth of volume and $p$ is the pressure.

Here one faces the first puzzle. As we have mentioned in the Introduction, according to the current observational data, the expansion
of the universe is accelerated and this is due to the existence of the negative pressure $(p = - \mid p \mid <0)$ matter or dark energy \cite{PhysToday2011}. In relation to this, the internal energy $dE$ of the universe {\it grows} proportionally to the growth of volume due to negative pressure. In fact, this negative pressure does positive work onto the system which is the universe, while the system does negative work (in analogy to what happens to a body falling in gravitational field of the Earth). On the contrary, positive pressure $p>0$ does negative work trying to shrink the universe while it is expanding $dV > 0$ (and the universe does positive work loosing its energy) - negative pressure acts in the ``right'' direction which means that it leads to the expansion of the universe and to the increase of its volume. It is worth emphasizing that the total energy is obviously conserved ($dS=0$ in (\ref{1stlaw})).

The second puzzle appears, when instead of internal energy one considers the energy density $\varepsilon$ (the mass density is $\varrho = \varepsilon/ c^2$). Since
\be
\label{endensity}
E = \varepsilon V~,
\ee
then taking the differentials one has
\be
V d \varepsilon + dV \varepsilon = - p dV~~,
\ee
which after simple rearrangements leads to
\be
\label{dero}
d\varepsilon = - (\varepsilon + p) \frac{dV}{V}~,
\ee
and so we have a very surprising result. If the universe expands ($dV > 0$, $V>0$), then {\it its energy density diminishes} $d\varepsilon <0 $ which
is quite obvious intuitively (similarly as the density of the condensed air revealed from a container). However, it happens only
{\it for positive and slightly negative} pressure which fulfills the condition (known as the null energy condition \cite{wald})
\be
\label{nec}
p > - \varepsilon
\ee
or
\be
p < 0 \hspace{1.cm} {\rm and} \hspace{1.cm} 0 > p > - \varepsilon~~.
\ee
Surprisingly, for the very negative pressure
\be
\label{nonec}
p < - \varepsilon~~,
\ee
when the volume grows ($dV >0$), the energy density also {\it grows} ($d\varepsilon > 0$), and this is exactly the effect of phantom which is the dark energy
of such a curious property. This effect leads to the accumulation of higher and higher energy {\it at every point} of the universe ($\varepsilon \to \infty$) as well as in the universe {\it as a whole} ($E \to \infty$). The singular state accompanied to these infinities was dubbed a big rip \cite{robert}, since it gives the effect that all the structures which were present in the universe (galaxies, planetary systems, atoms etc.) are destroyed (``ripped'') \cite{CCW}. It is in a way an opposite state to a big crunch (the final state for a recollapsing universe) in which all the structures become ``crushed'' to a zero volume \cite{PRD03}.

\section{Standard particle and phantom collision puzzles}
\label{collision}
\setcounter{equation}{0}

We assume the kinetic energy and momentum conservation and consider an elastic collision of a standard particle of mass $m$, which possesses positive kinetic
energy and positive momentum with phantom of mass $m_f$, which possesses negative kinetic energy and momentum. We will derive the formulas for the velocities of these
particles after collision and try to prove that due to its negative energy, phantom can transfer the energy to the standard particle leading to the growth of its velocity
and so its energy.

Using the standard textbook formulas for an elastic collision \cite{young} of a standard particle with phantom and bearing in mind that phantom has negative kinetic
energy and momentum we may write
\bea
\label{Ek}
\frac{1}{2} m v_1^2 - \frac{1}{2} m_f v_{f1}^2 &=& \frac{1}{2} m v_2^2 - \frac{1}{2} m_f v_{f2}^2~~,\\
\label{ped}
m v_1 - m_f v_{f1} &=& m v_2 - m_f v_{f2}~~,
\eea
where $m$ - mass of the standard particle, $m_f$ - mass of phantom, $v_1$ - velocity of the standard particle before collision, $v_2$ - velocity of the standard particle
after collision, $v_{f1}$ - velocity of phantom before collision, $v_{f2}$ - velocity of phantom after collision. Following the textbook derivation of the final velocity
formulas one can rewrite (\ref{Ek}) and (\ref{ped}) in the form:
\bea
\label{separ1}
m_1 (v_1 - v_2)(v_1 + v_2) &=& m_f (v_{f1} - v_{f2})(v_{f1} + v_{f2})~~,\\
\label{separ2}
m (v_1 - v_2) &=& m_f (v_{f1} - v_{f2})~~.
\eea
Using (\ref{separ2}), the Eq. (\ref{separ1}) gives
\be
\label{suma}
v_{f2} = v_1 + v_2 - v_{f1}~~,
\ee
which after substituting into (\ref{separ2}) allows to write
\be
\label{zmiana}
m (v_1 - v_2) = m_f (2v_{f1} - v_1 - v_2)~~,
\ee
and after some manipulations we have
\be
\label{v2}
v_2 = \frac{2m_f}{m_f - m} v_{f1} - \frac{m_f + m}{m_f - m} v_1~~.
\ee
Substitution of (\ref{zmiana}) into (\ref{suma}) gives
\be
\label{vf2}
v_{f2} = \frac{m_f + m}{m_f - m} v_{f1} - \frac{2m}{m_f - m} v_1~~.
\ee
There is a big difference between textbook formulas for collisions of standard particles and the formulas (\ref{v2}) and (\ref{vf2})
since they {\it cannot} be applied for the collision of a standard particle and phantom of equal masses $m_f = m$, unless both particles are initially at rest
i.e. $v_1 = v_{f1}$ (formally $v_2 \to \infty$ and $v_{2f} \to \infty$, when $m_f \to m$). This case will be considered separately. However, before we deal with such a case, we will first discuss how the colliding particles behave if one of them has much larger mass than another. For example, if phantom mass is negligible
$m_f \ll m$ $(m_f \approx 0)$, then
\be
\label{mfmale}
v_1 = v_2 \hspace{0.5cm} {\rm and} \hspace{0.5cm}v_{f2} = 2 v_2 - v_{f1}~~,
\ee
while if a standard particle has negligible mass $m \ll m_f$ $(m \approx 0)$, then
\be
\label{mmale}
v_{f2} = v_{f1} \hspace{0.5cm} {\rm and} \hspace{0.5cm} v_2 = 2 v_{f1} - v_1~~.
\ee
It means that the heavier particle practically moves without any change of its velocity, while a lighter particle has final velocity opposite to its initial value plus the doubled velocity of a heavier particle.

Let us assume that an ordinary particle is at rest initially, i.e. that $v_1 = 0$ (which is just a scattering of a phantom on a standard particle).
Then, from (\ref{Ek}) we obtain that
\be
\frac{1}{2} m_f \left( v_{f2}^2 - v_{f1}^2 \right) = \frac{1}{2} m v_2^2~~,
\ee
so that one has:
\be
v_{f2}^2 > v_{f1}^2~~,
\ee
and
\be
\label{warunek}
- \frac{1}{2} m_f v_{f2}^2 < - \frac{1}{2} m_f v_{f1}^2~~,
\ee
which means that the (negative) kinetic energy of phantom after collision {\it is smaller} than
(negative) kinetic energy of phantom before collision. In other words, the standard particle {\it explores} energy from phantom.

Another case is when phantom is initially at rest, i.e. when $v_{f1} = 0$. In such a case the Eq. (\ref{Ek}) gives
\be
\frac{1}{2} m \left( v_1^2 - v_2^2 \right) = - \frac{1}{2} m_f v_{f2}^2~~,
\ee
which after bearing in mind negativity of the phantom energy requires that
$v_2^2 > v_1^2$. This means that the standard particle, which is incident onto (initially at rest) phantom of zero energy,
gains energy from phantom and starts moving faster that it did, while phantom ``falls'' into a state of lower (negative) energy.
In a general case the situation is similar, since from (\ref{Ek}) we have
\be
\frac{1}{2} m \left( v_1^2 - v_2^2 \right) = - \frac{1}{2} m_f \left( v_{f2}^2 - v_{f1}^2 \right)
\ee
and because $v_{f2}^2 > v_{f1}^2$ (so that the condition (\ref{warunek}) is fulfilled), then also
$v_2^2 > v_1^2$, which means that phantom ``falls'' into a lower (negative) kinetic energy state
and the incoming standard particle enlarges its velocity and its energy.

Finally, let us consider the case when the masses of both colliding particles (standard and phantom) are equal $m=m_f$. We apply formulas
(\ref{Ek})-(\ref{ped}) and assume that the standard particle is initially at rest, i.e.
\bea
-\frac{1}{2} m v_{f1}^2 &=& - \frac{1}{2} m v_{f2}^2 + \frac{1}{2} m v_2^2~~,\\
- m v_{f1} &=& - m v_{f2} + m v_2~~,
\eea
which after some manipulations give
\bea
v_2^2 &=& (v_{f2} - v_{f1})(v_{f2} + v_{f1})~~,\\
v_2 &=& v_{f2} - v_{f1}~~,
\eea
and so
\bea
\label{vr1}
v_2 &=& v_{f2} + v_{f1}~~,\\
\label{vr2}
v_2 &=& v_{f2} - v_{f1}~~.
\eea
The Eqs. (\ref{vr1})-(\ref{vr2}) are consistent, provided that
\be
v_2 = v_{f2} \hspace{0.2cm} {\rm and} \hspace{0.2cm} v_{f1}= 0~~.
\ee
Physical interpretation is in order. Namely, we deal with a kind of a (classical) pair creation here - the standard particle and the phantom
with opposite values of momentum and energy (giving totally zero) appear simultaneously \cite{perkins}. This is consistent, since we have assumed validity of
the conservation of energy and momentum at the collision.

Exactly the same result is obtained if we assume that initially phantom is at rest, i.e.
\bea
\frac{1}{2} m v_1^2 &=& \frac{1}{2} m v_2^2 - \frac{1}{2} m v_{f2}^2~~,\\
m v_1 &=& m v_2 - m v_{f2}~~,
\eea
or
\bea
(v_1 - v_2)(v_1 + v_2) &=& - v_{f2}^2~~,\\
v_1 - v_2 &=& - v_{f2}~~,
\eea
so that
\bea
v_1 + v_2 &=& v_{f2}~~,\\
v_1 - v_2 &=& - v_{f2}~~,
\eea
which can be fulfilled only if
\be
v_1 = 0 \hspace{0.5cm} {\rm oraz} \hspace{0.5cm} v_2 = v_{f2}~~.
\ee

\section{Summary}
\label{summary}
\setcounter{equation}{0}

Cosmology and physics is full of puzzles. Very often it appears that in order to explain the observational data, we have to apply some completely new ideas which earlier
sounded an absurd for us. This is exactly what happened with phantom - a kind of not yet observed type of matter which effect on the evolution of the universe is only gravitational. This matter characterizes itself by the very negative pressure which in general relativity gives the effect of antigravity and leads to some non-standard phenomena which we have described in this article. It has been shown that in order to get stable phantom oscillations one would have to use an ``upside down'' Newtonian potential. Besides, phantom leads to a growth of the energy density in the universe in accompany to its expansion (matter coagulates rather than dilutes), and also it leads to the effect of loosing its energy to scattered particles due to negative kinetic energy which is lowered at the collision. However, the most surprising result in the context of standard Newtonian physics which, as far as we are aware, was not yet discussed in the literature, is the fact that the only way to collide phantom particle and the standard particle of the same mass in a conserved classical way is the creation of these particles as a pair with opposite values of momenta and kinetic energy. This is an amazing classical analogue of quantum mechanical phenomenon of a pair creation (e.g. an electron and a positron) in a strong electric field or in quantum vacuum.

\section{Acknowledgements}

I would like to thank Robert R. Caldwell and Paul C.W. Davies for enlightening discussions. The support from the Polish National Science Center grant
DEC-2012/06/A/ST2/00395 is also acknowledged.
%This work was greatly advanced while visiting Beyond Center for Fundamental Concepts in Science, Arizona State University.

%%%%%%%%%%%%%%%%%%%%%%%%%%%%%%%%%%%%%%%%%%%%%%%%%%%%%%%%%%%%%%%%%%%%%
%%%%%%%%%%%%%%%%%%%%%%%%%%%%%%%%%%%%%%%%%%%%%%%%%%%%%%%%%%%%%%%%%%%%%

\end{document}